\documentclass[aps,prb,reprint,superscriptaddress,showpacs]{revtex4-1}

\usepackage[]{graphicx}
\usepackage{epsfig}
\usepackage{color}

\newcommand{\tn}{t_{\text{nom}}}

\begin{document}

\title{Kondo screening and beyond: an x-ray absorption and dichroism study of CePt$_5$/Pt(111) 
}



\author{C.~Praetorius}
\affiliation{Physikalisches Institut, Universit{\"{a}}t W{\"{u}}rzburg, Am Hubland, 97074 W{\"{u}}rzburg, Germany}

\author{K.~Fauth}
\email[]{fauth@physik.uni-wuerzburg.de}
\affiliation{Physikalisches Institut, Universit{\"{a}}t W{\"{u}}rzburg, Am Hubland, 97074 W{\"{u}}rzburg, Germany}
\affiliation{Wilhelm Conrad R{\"{o}}ntgen-Center for Complex Material Systems (RCCM), Universit{\"{a}}t W{\"{u}}rzburg, Am Hubland, 97074 W{\"{u}}rzburg, Germany}

\date{\today}

\begin{abstract}
We use x-ray absorption spectroscopy as well as its linear and circular magnetic dichroisms to characterize relevant interactions and energy scales in the surface intermetallic CePt$_5$/Pt(111).
The experiments provide insight into crystal field splitting, effective paramagnetic moments, their Kondo screening and mutual interactions and thus into many aspects which typically determine the low temperature behavior of correlated rare earth compounds.
Exploiting the tuneability of Ce valence through the thickness dependent epitaxial strain at the CePt$_5$/Pt(111) interface, we are able to systematically investigate the impact of hybridization strength on these interactions. 
Considerable Kondo screening is indeed observed at all CePt$_5$ thicknesses, and found to be strongest in case of strongest hybridization.
While the magnetic response is commensurate with an impurity Kondo scale of $T_K \gtrsim 10^2$ K for specimen temperatures $T \gtrsim 30$ K, this is no longer the case at lower temperature.
Its detailed study by XMCD at one specific thickness of CePt$_5$ reveals an anomaly of the susceptibility at $T^* \approx 25$ K instead, which we tentatively associate with the onset of lattice coherence.
At lowest temperature we observe paramagnetic saturation with a small Ce $4f$ saturation magnetization.
Within the framework of itinerant $4f$ electrons, saturation is due to a field induced Lifshitz transition involving a very heavy band with correspondingly small degeneracy temperature of $T_F \approx 7$ K. 
This small energy scale results in the persistence of Curie-Weiss behavior across the entire range of experimentally accessible temperatures ($T \gtrsim 2$ K). 
Our work highlights the potential of magnetic circular dichroism studies in particular for Kondo and heavy fermion materials, which so far has remained largely unexplored.
\end{abstract}


\maketitle

\section{Introduction}

The richness and complexity of physical behavior encountered in Ce intermetallics derives from the interaction of localized and itinerant electronic degrees of freedom, i.~e.~the finite hybridization of Ce $4f$ states with the band structure of the periodic solid.
The microscopic details of the interactions give rise to a rich phenomenology of physical properties and a variety of ground states including magnetic order, superconductivity and paramagnetic heavy fermion liquids
\cite{Hews93a,Grew91a,loeh07a,gege08a,yang08a,steg16a,yang16a}. 
This variability arises from the occurrence of competing effective interactions with small associated energy scales.
Accordingly, tuning the interactions by nonthermal control parameters such as hydrostatic or chemical pressure may result in quantum critical points and unconventional behavior in their vicinity
\cite{gege08a,yang08a,klei08a,si10a,zhon13a,varm16a}.

Identifying and characterizing the relevant energy scales thus constitutes an essential part of understanding the low temperature behavior and of establishing correlations such as e.~g.~between local hybridization strength on the one hand and macroscopic properties on the other. 
In this respect, advanced methods of surface science have demonstrated tremendous potential and novel insight in recent years, notably owing to their resolving capabilities in real or reciprocal space in combination with great spectral resolution.
Associating findings from surface sensitive experiments to bulk properties of the respective materials may represent a nontrivial task, since relevant interactions are frequently altered in the vicinity of the surface  \cite{erns11a,hami11a,iwam95a,Dall02a,guet14a,mula14a,Pati16a}.

Technical limitations in applying nonthermal control parameters as well as the unavailability of classical thermodynamic methods or inelastic neutron scattering restrict the possibilities of systematically studying the properties and phase diagrams of systems at surfaces in an analogous manner to bulk materials.
In the present work, we overcome some of these limitations by exploiting the fact that epitaxial strain at an interface may serve as a parameter controlling the strength of hybrizidation between Ce $4f$ states and those of the metallic bands \cite{Prae15a}. 
X-ray circular magnetic dichroism (XMCD) is then being used as an element and orbital specific probe of the anisotropic Ce $4f$ paramagnetic response in these ultrathin specimens including its temperature dependence.
In this way, we systematically tune the many body interactions via Ce $4f$ hybridization and study its relevance for various electronic and magnetic properties such as crystal field splitting and magnetic Kondo screening.

\begin{figure} 
\includegraphics[width=\columnwidth]{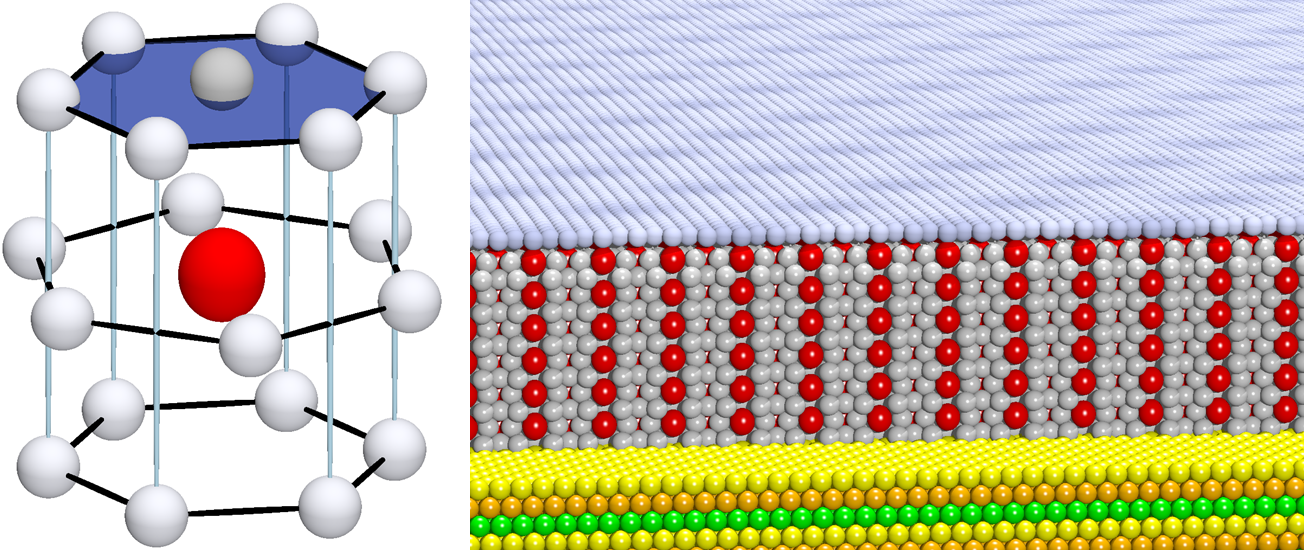} 
\caption{\label{fig1struct}
(\textit{left}) Hexagonal atomic environment of a Ce atom (large red sphere) in CePt$_5$. 
The lower layer exhibits the kagome hole characteristic of the bulk lattice.
The top layer is shown with an additional Pt atom at this position, representing the surface termination \cite{Tere15a,Prae15b}.
(\textit{right}) Atomic arrangement of an idealized CePt$_5$/Pt(111) specimen at $\tn = 6$ u.c. (structure C' in Refs.~\onlinecite{Kemm14a,Prae15a}, which is the majority phase at this thickness).
The periodic brightness modulation of the surface atoms corresponds to the superstructure corrugation observed in scanning tunneling microscopy \cite{Kemm14a}.
}
\end{figure}

Among the ordered binary bulk intermetallic phases of Ce and Pt, CePt$_5$  is the one richest in Pt \cite{Pred93a,Jang10a}.
It crystallizes in the hexagonal CaCu$_5$ structure  (space group P6/mmm, No.~191).
The local point group at the Ce sites is D$_{6h}$.
Previous work has established that alloying Ce into the surface of Pt(111) results in surface intermetallics which adopt the same atomic lattice \cite{Badd97a,Esse09a,Kemm14a,Prae15a},
except for the surface termination \cite{Tere15a,Prae15b} where a dense Pt atomic layer is formed by occupying the kagome hole positions with extra Pt atoms, as shown in Fig.~\ref{fig1struct}.
Careful preparation results in well defined CePt$_5$ thickness at the specimen surface, which we refer to as the nominal thickness $\tn$ in multiples of the CePt$_5$ unit cell (u.c.) along the hexagonal axis.
The intermetallic thickness amounts to approx.~$0.44$ nm per u.c. of CePt$_5$.

Early measurements of the bulk magnetic response in polycristalline CePt$_5$ were interpreted within a crystal field scheme involving a fairly large overall Ce $4f$ level splitting of about $76$ meV \cite{Luek79a}.
Later work \cite{Schr88a} concluded on antiferromagnetic ordering  at $T_N=1$ K based on low temperature susceptibility and specific heat measurements.
Overall, the thermodynamic data gave no evidence for particular importance of Kondo or heavy fermion physics at the time.
A resistivity minimum at $T\lesssim 10$ K, found lateron by  Sagmeister et al.~\cite{Sagm97a} might hint at a Kondo scale of that order, the marked resistivity decrease below $T=2$ K was linked to magnetic ordering rather than coherent band formation, however.

Matters seem different with the CePt$_5$/Pt(111) surface intermetallics.
Temperature effects in angle resolved photoemission were first observed by Andrews et al.~\cite{andr95a}
and lateron identified as the tail of a Kondo resonance by Garnier et al.~\cite{Garn97a} in specimens which we identify as CePt$_5$ with  $\tn = 4\dots5$ u.c. based on their electron diffractograms \cite{Kemm14a,Prae15a}.
A re-investigation by Klein et al.~\cite{Klei11a} demonstrated the persistence of the Kondo resonance to $T=66$ K and signatures of incipient coherence at $T\approx 13$ K.
Thin film CePt$_5$ thus appears to be a Kondo lattice material with $T^* < T_K$, the opposite scenario compared to a class of heavy fermion materials for which a phenomenological two-fluid picture has been proposed in the last years 
\cite{yang08a,shir12a,yang14a,jian14a,yang16a}.

Hints at the relevance of Kondo physics in CePt$_5$/Pt(111) were also detected recently by x-ray absorption (XA) and magnetic circular dichroism (XMCD) experiments \cite{Prae15a}.
The temperature dependence of the Ce valence measured by XA hinted at a Kondo scale in excess of $10^2$ K, lending support to the $T^* < T_K$ scenario.
CePt$_5$/Pt(111) also displays a remarkable dependence of the Ce valence as a function of intermetallic thickness, resulting in an interesting tunability of the electronic and magnetic properties.
The CePt$_5$ thickness may thus be used as a non-thermal control parameter for the interactions in this material, while the underlying atomic structure is essentially unchanged\cite{Prae15a}.

In our preliminary analysis of a restricted XMCD dataset~\cite{Prae15a}, we were also able to show that a considerable degree of magnetic Kondo screening must be present in these CePt$_5$ thin film specimens.
Here, we extend over those results by examining the x-ray linear dichroism (XLD) and the anisotropic Ce $4f$ magnetic response detected by XMCD over a larger temperature range.
These experiments yield valuable information on crystal field splitting \cite{Cast95a,Hans08a,Will10a,Will11a,Prae16a}, Kondo screening \cite{Prae15a} and magnetic coupling \cite{Prae16a} in the specimens and we shall discuss in detail how we obtain these quantities from our experimental data.
Moreover, we report on a low temperature anomaly in the (inverse) susceptibility, which we tentatively interpret as an independent signature of a coherence scale of $T^* \approx 25$ K.
X-ray absorption experiments may thus serve as a powerful means to identify the various interaction scales which all contribute to the complexity in the behavior of Kondo lattice materials.

\section{methods}

CePt$_5$/Pt(111) specimens were produced by following the procedures described in our previous work\cite{Kemm14a,Prae15a}.
Clean Pt(111) was prepared by repeated cycles of 1 keV Ar$^+$ ion sputtering and annealing to 1170 K.
Cerium (99.9\% purity) was evaporated onto this surface near ambient temperature and interdiffusion was activated by subsequent annealing to approx.~970 K for 5 to 10 min.
This procedure results in well-ordered CePt$_5$ intermetallic phases the thickness of which is predetermined by the quantity of Ce deposited \cite{Kemm14a,Prae15a}.
They are terminated a by a single dense Pt(111) atomic layer, giving rise to the remarkable inertness of these surfaces \cite{Prae15b}.

Soft x-ray Ce M$_{4,5}$ XAS and XMCD experiments were carried out at the PM 3 bending magnet beam line for circular polarization of BESSY II at Helmholtz Center Berlin (HZB).
Absorption spectra were acquired in the total electron yield mode (TEY) using circular polarized radiation (polarization: $\approx 0.93$) within a custom XMCD end station ($\pm 3$ T superconducting UHV magnet).
Appropriate normalization is achieved by simultaneous measurement of the TEY from a gold mesh.
Owing to the polarization characteristics of the beam line and the experiment geometry, spectra taken at an angle of x-ray incidence of $\theta_X=60^\circ$ with respect to the surface normal probe the polarization-averaged, ``isotropic'' spectrum \cite{Prae16a}. 

Additional datasets were acquired at the SOLEIL SR facility using the local CroMag end station at the DEIMOS beam line \cite{ohre14a}.
Measurements were taken with better energy resolution, a degree of circular polarization near $100$\% and with the main goal of reaching lower specimen temperatures.
The same experimental geometries were chosen to warrant a maximum of comparability of the XMCD data.
Based on polarization alone, this choice entails the linear dichroism between normal and oblique incidence data to be enhanced by $12.5$\% with respect to to PM3 under otherwise identical settings.

The TEY escape depth was previously found to be of the order of 1-$1.5$ nm \cite{Prae15a}.
In combination with the moderate concentration of Ce in the CePt$_5$ specimens, this creates a situation in which TEY saturation effects \cite{naka99a} may safely be neglected. 

In our analyses we make use of simulated absorption spectra, obtained from full atomic multiplet calculations as implemented in the Quanty package \cite{Have12a,quantylink}.
The parameters of the calculations are similar to those employed in our preceding analysis of CeAg$_x$ \cite{Prae16a}.
\textit{ff (df)} Slater Integrals were reduced to 60\%  (80\%) 
of their respective Hartree-Fock values and energy dependent core excitation lifetimes were introduced to achieve best agreement with the experimental line shapes.

\section {results and discussion}
\subsection{general observations and discussion framework} \label{general}

\begin{figure} 
\includegraphics[width=\columnwidth]{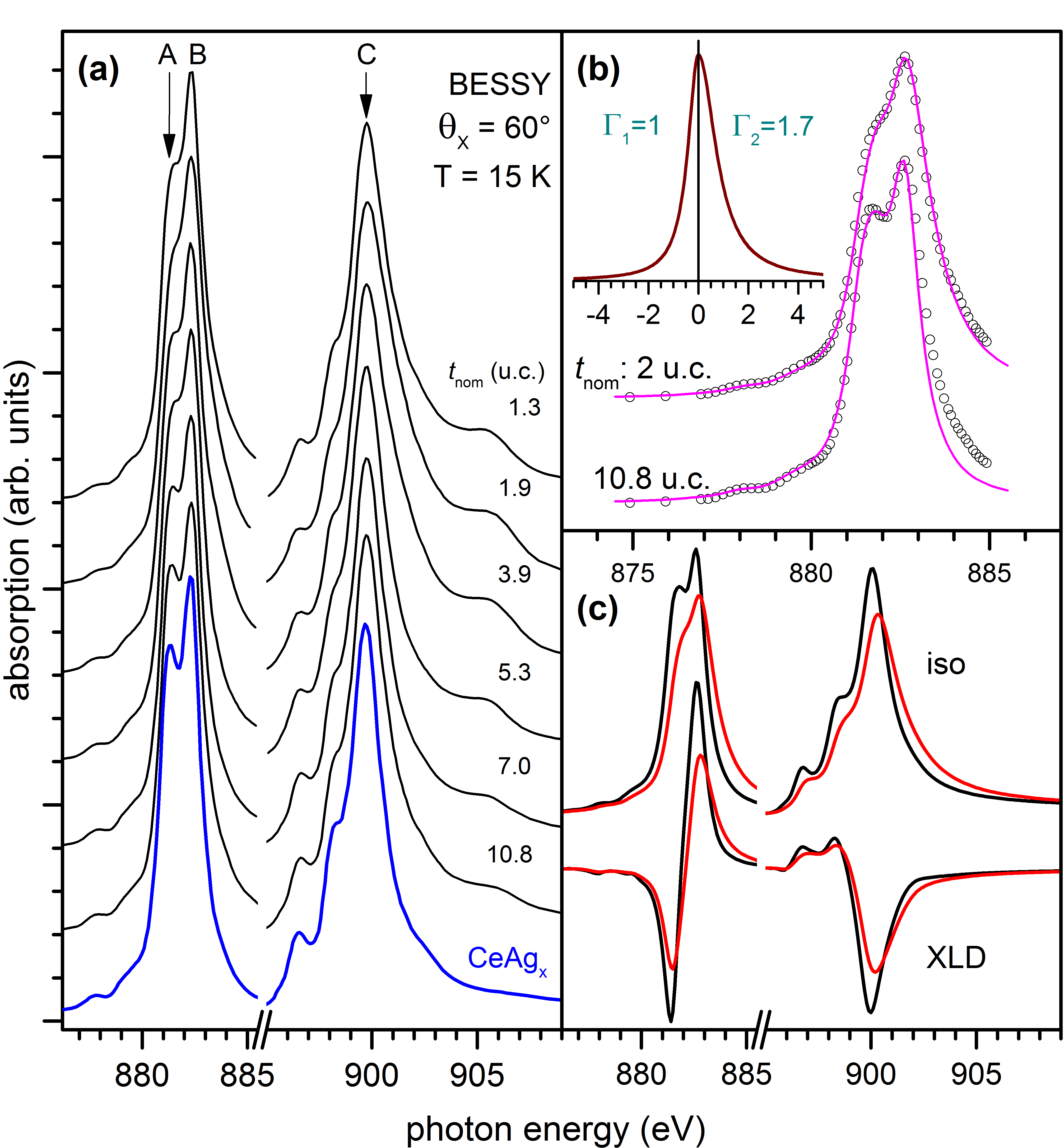} 
\caption{\label{fig2XASXLD}
(a) Isotropic Ce M$_{4,5}$ XA spectra as a function of CePt$_5$ thickness along with a similar spectrum acquired on CeAg$_x$ \cite{Prae16a}. 
Most prominent peaks are labeled as A,B, and C. Peak A reduces to a mere shoulder, when the high energy shoulder at 906 eV is strongest ($\tn \approx 2$ u.c.)
(b) Comparison of the M$_5$ spectra (symbols) and analysis of their line shape. For $\tn=10.8$ u.c., the solid line is given by a simulated Ce M$_5$ spectrum computed using Quanty \cite{Have12a,quantylink}.
For $\tn=2$ u.c., the simulated spectrum has been convoluted with the asymmetric response function shown in the inset.
(c) calculated effect of the asymmetric response on XA and XLD line shapes and magnitudes. Note that relative to the XA strength, the XLD is most strongly reduced at feature B.
}
\end{figure}

Figure~\ref{fig2XASXLD}(a) displays a selection of Ce M$_{4,5}$ spectra from CePt$_5$ specimens with different thicknesses alongside with data acquired on a similar CeAg$_x$ sample \cite{Prae16a}.
All datasets were obtained at oblique incidence (PM3 beamline) and hence represent the isotropic spectra \cite{Prae16a}.
In contrast to the CeAg$_x$ XA spectrum, a high energy shoulder appears in the CePt$_5$ data at $\hbar\omega \approx 906$ eV.
Its presence is indicative of finite hybridization between the localized Ce $4f$ states and the metallic band structure \cite{Gunn87a} and hence an admixture of states with $n_{4f} = 0$ character.
Analyzing the spectral intensity of this shoulder reveals a non-monotonic dependence of the Ce valence on the thickness of the intermetallic film \cite{Prae15a}.
Interestingly, there is a systematic concommitant variation in the line shape of the main $f^1 \rightarrow d^9f^2$ type excitation spectrum, most easily visible as the variation of relative spectral weight of feature A in Fig.~\ref{fig2XASXLD}(a).
The greatest similarity with the CeAg$_x$ spectrum is given in case of weakest hybridization, i.e.~for $\tn = 10.8$~u.c.
The detailed analysis of the XA line shape \cite{CPdiss} reveals that the main contribution to its variation is an increasingly asymmetric response of the individual resonances as the hybridization is increased.

This is illustrated in Fig.~\ref{fig2XASXLD}(b) by a comparison of the isotropic spectra of weakest and strongest hybridization, i.e. at $t_{\text{nom}} = 10.8$ u.c. and $t_{\text{nom}} = 2$ u.c., respectively.
Along with each experimental M$_5$ spectrum a simulation is being displayed.
The simulation for  $t_{\text{nom}} = 10.8$ u.c. uses symmetric Lorentzians ($\Gamma=1$ eV) to represent the lifetime of the core hole excitation, while asymmetric ones (see inset) are employed for $\tn = 2$ u.c.
The pronounced asymmetric spectral response causes the (apparent) reduction of the spectral weight of feature A vs.~feature B, at the same time the falling edge of the M$_5$ absorption is significantly broadened.
The strength of asymmetry in the spectral response correlates directly with the $f^0$ spectral weight \cite{Prae16NNa}.
We therefore assume that it is a manifestation of an increasingly efficient coupling to low energy excitations in the metallic bands with which the $4f$ states hybridize and as such related to the edge singularity problem (see e.g. Refs.~\onlinecite{mauc09a,Sipr11a,vins11a,have14a} and references therein).
Its study for quasi-atomic multiplets coupled to a metal is as a challenging problem \cite{mauc09a,have14a}.
Our hypothesis therefore is currently under detailed theoretical scrutiny \cite{Prae16NNa}.

The analysis in Refs.~\onlinecite{CPdiss,Prae16NNa} suggests in particular that hybridization induced ground state admixture of states with $j=7/2$ character, clearly identified e.g.~in CeFe$_2$ \cite{Delo99a}, may safely be neglected in the present case.
Likewise, the neglect of double occupancy of the $4f$ orbital as a result of strong on-site Coulomb repulsion is a good approximation for CePt$_5$.

Nevertheless, the redistribution of spectral weight towards higher excitation energies not only affects the spectral appearance of the XA spectra but also modifies the spectral shape of XLD.
This is shown in Fig.~\ref{fig2XASXLD}(c), where we plot simulated XLD spectra with the same parameters as in Fig.~\ref{fig2XASXLD}(b) along with their isotropic XA counterparts.
While the peaks of both XA and XLD spectra are reduced in intensity, we note that the ratios between the magnitudes of XLD and XA are affected in different ways for the main spectral features.
In particular, while the loss of peak intensity of feature B in XA is partly compensated for by transfer of spectral weight from feature A, the same spectral weight transfer additionally reduces the XLD of feature B owing to the sign change in XLD between features A and B.
Overall, the observation of an asymmetric spectral response leads us to expect a systematic decrease of the relative XLD amplitude as the hybridization is increased.

Experimentally, we shall find this expectation confirmed (see below), but the magnitude of XLD is even suppressed well beyond the level to be expected by the modified $f^1$ XA line shape.
Similar findings have been reported before and have been suggested to result from hybridization viz.~the Kondo interaction \cite{Will10a}.
Within the non-crossing approximation \cite{Bick87b} (NCA) to the impurity problem this may qualitatively be rationalized as follows.
We represent the many body state as a superposition of states with $f^0$ and $f^1$ character

\begin{equation}\label{inistate}
\left | \Psi \right \rangle = c_0 \left | f^0 \right \rangle + c_1 \left | f^1 \right \rangle  = c_0(T) \left | f^0 \right \rangle + \sum_{m_j} c_{m_j}(B,T) |m_j\rangle.
\end{equation}

In the absence of excited state mixing, the squared coefficients $c_0^2$ and $c_1^2$  directly correspond to the relative spectral weights of the $f^0$ and $f^1$ related fractions of the XA spectra \cite{Jo87a,Gunn87a}.
In a second step we rewrite the $f^1$ part as a superposition of the six $|m_j\rangle$ states of the Ce $4f$ $j=5/2$ multiplet.
In the NCA treatment, each of these states is represented by the resonance of a temperature dependent many body spectral function.
Resonance positions are given by the CF scheme and Zeeman energies, and they acquire finite width through hybridization.
The statistical weight of each level is obtained by integration over these spectral functions. 
In particular, due to the hybridization induced width the excited states possess higher occupation probabilities at low temperature than what would be obtained from ordinary statistics of discrete CF states.
As a result, the magnitude of XLD is reduced.

Just as the XLD, the magnetic response is mainly governed by the occupation of the CF states.
Zwicknagl et al.~\cite{Zwic90a} have proposed a simplified NCA scheme which is particularly suited to treat CF splitting and mean field coupling at the same level as the many body Kondo physics.
The characteristic failure of the standard NCA at low temperature is circumvented in this approach by omission of the divergent term in the spectral function of the $|f^0\rangle$ state.
The application of this scheme does not yield a satisfactory description of our experimental results though:
imposing a Kondo scale of order $10^2$ K in accordance  with the observed Ce $4f$ occupation $n_{4f}(T)$\cite{Prae15a} leads to a much stronger supression of the low temperature magnetic response than we observe.
Conversely, assuming a small effective Kondo scale ($T_K \lesssim T^*$) in the NCA, one computes much too large a susceptibility at higher temperatures compared with the experimental data.
We therefore reckon that it is the occurrence of two distinct energy scales which prevents the applicability of the NCA scheme in case of CePt$_5$/Pt(111).

\subsection{XLD analysis} \label{ResXLD}

\begin{figure} 
\includegraphics[width=\columnwidth]{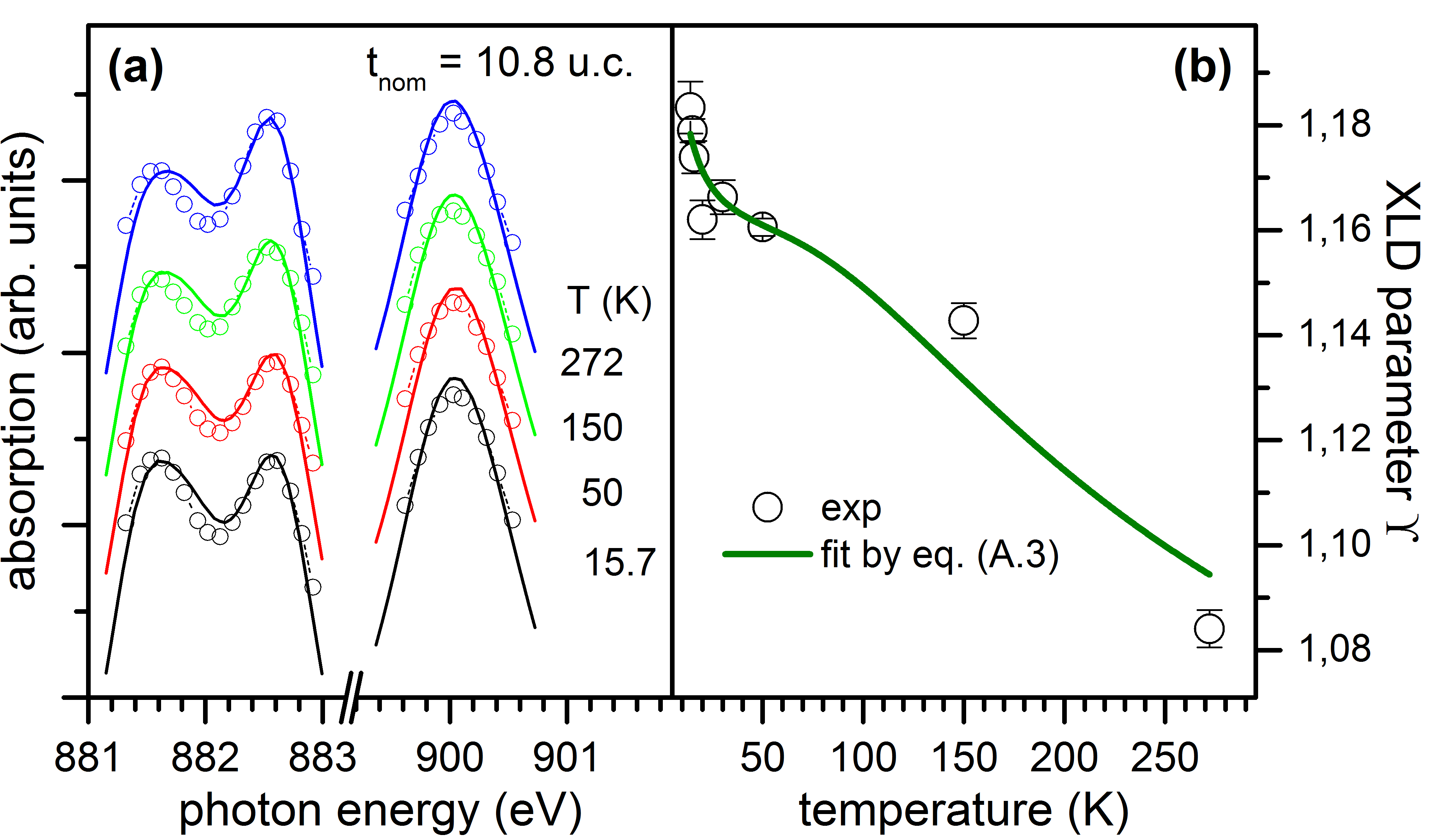} 
\caption{\label{fig3XLD11uc}
(a) Symbols: selection of normal incidence spectra for $\tn=10.8$ u.c. in the vicinity of their most prominent peaks, recorded at various temperatures.
Solid lines: fits by superpositions of simulated absorption spectra according to eq.~(\ref{NImodel}).
(b) Symbols: XLD parameter $\Upsilon(T)$, extracted from the experimental spectra according to eq.~(\ref{DefUps}).
Solid line: fit to $\Upsilon(T)$ using eq.~(\ref{EqUps1}).
}
\end{figure}

In our analysis below, we shall therefore use a conventional crystal field approach and take the many body aspects of the problem into account in a qualitative, phenomenological way.
In extension of our previous work \cite{Prae16a} we represent the possibility of non-thermal CF state occupation by introducing an `isotropic fraction'  of weight  $w_i$ in the normal incidence (NI) spectra, where for simplicity we assume this fraction to be
temperature independent and the same in the three Kramers doublets.
The model equation for the temperature dependence of NI spectra ($f^1$ part only) thus reads as follows:

\begin{equation} \label{NImodel}
I^{NI}(T) = \frac{1}{Z'} \left ( w_i I^{ISO} + \overline{w}_i  (  I^{NI}_{|1/2\rangle} + p_1 I^{NI}_{|3/2\rangle} + p_2 I^{NI}_{|5/2\rangle}  )   \right ).
\end{equation}

Here,  $ \overline{w}_i=(1-w_i)$, and $p_{1,2} = \exp (-\Delta_{1,2}/k_BT)$ are the Boltzmann weights representing the thermal excitation probabilities
according to the CF splittings $\Delta_1= E_{3/2}-E_{1/2}$ and  $\Delta_2= E_{5/2}-E_{1/2}$. The standard partition function $Z$ being given by $Z=  1 + p_1 + p_2 $, $Z'$ is constructed such as to take
the isotropic fraction into account, i.e. $Z' = w_i  + \overline{w}_i Z$.

Figure~\ref{fig3XLD11uc}(a) displays a selection of NI spectra from the CePt$_5$ specimen with smallest hybridization ($\tn = 10.8$ u.c.) along with the fits
according to eq.~(\ref{NImodel}).
The best fit is obtained for $\Delta_1 = 1.8 \pm 1.5$ meV, $\Delta_2 = 27 \pm 6$ meV and yields $w_i \approx 0.23$.
We thus find a similar energetic ordering of the CF states as in CeAg$_x$ while the total CF splitting (i.e. $\Delta_2$) is larger by about one order of magnitude in CePt$_5$.

While the data presented in Fig.~\ref{fig3XLD11uc}(a) demonstrate that the fits do adequately capture the essential thermal evolution of the XA spectra at normal incidence,
the fact that the simulated spectra do not perfectly match the experimental peak positons and line shape gives rise to a non-negligible residual error and thus a fairly shallow optimum with associated parameter uncertainties that are relatively large.
The situation further aggravates as the CePt$_5$ layer thickness is reduced owing to both the reduction Ce M$_{4,5}$ TEY signal above background and the increasing hybridization which reduces the magnitude of XLD as discussed above.
We have therefore sought an alternative means to evaluate the CF splitting from the temperature dependent XA data in a way that does not depend on an accurate simulation of the XA line shape.
A simple yet robust measure of the magnitude of XLD can indeed be found.
It essentially consists of a relation involving ratios of the peak amplitudes of features B and C in the normal and oblique incidence spectra, respectively (see Appendix for details).
The resulting XLD parameter $\Upsilon$ for $\tn = 10.8$ u.c. is plotted in Fig.~\ref{fig3XLD11uc}(b) as a function of specimen temperature.
From the definition of $\Upsilon$, an expression 
(eq.~(\ref{EqUps1}) in the appendix)
describing its temperature dependence is readily derived and the solid line in Fig.~\ref{fig3XLD11uc}(b) represents the best fit according to this model equation.
The essential behavior of the experimental data is well covered by the model, and we obtain $\Delta_1 = 0.6 \pm 0.3$ eV and $\Delta_2 = 29 \pm 4$ eV, largely in line with the foregoing analysis.
Unlike the fitting of entire spectra, however, the analysis in terms of $\Upsilon$ is readily carried out over the entire thickness range relevant to this study.
We note that just like in Fig.~\ref{fig3XLD11uc}(b) we observe $\Upsilon >1$ in all specimens and at all temperatures.
This finding immediately reveals that the CF state with  $|\pm 5/2\rangle$ character must be the one of highest energy.
The smaller energy scale $\Delta_1$ is responsible for the increase of $\Upsilon$ at low temperature in Fig.~\ref{fig3XLD11uc}(b).
For specimens with smaller $\tn$ such a low temperature variation of $\Upsilon$ could not unambiguously be determined.
We are therefore led to conclude that $\Delta_1$ assumes such small values that its temperature effect essentially slips out of the temperature range accessible in our XLD experiments.
We shall find this idea to be confirmed by the analysis of our XMCD data, the basics of which we discuss next.

\subsection{anisotropic paramagnetic response}

\begin{figure} 
\includegraphics[width=\columnwidth]{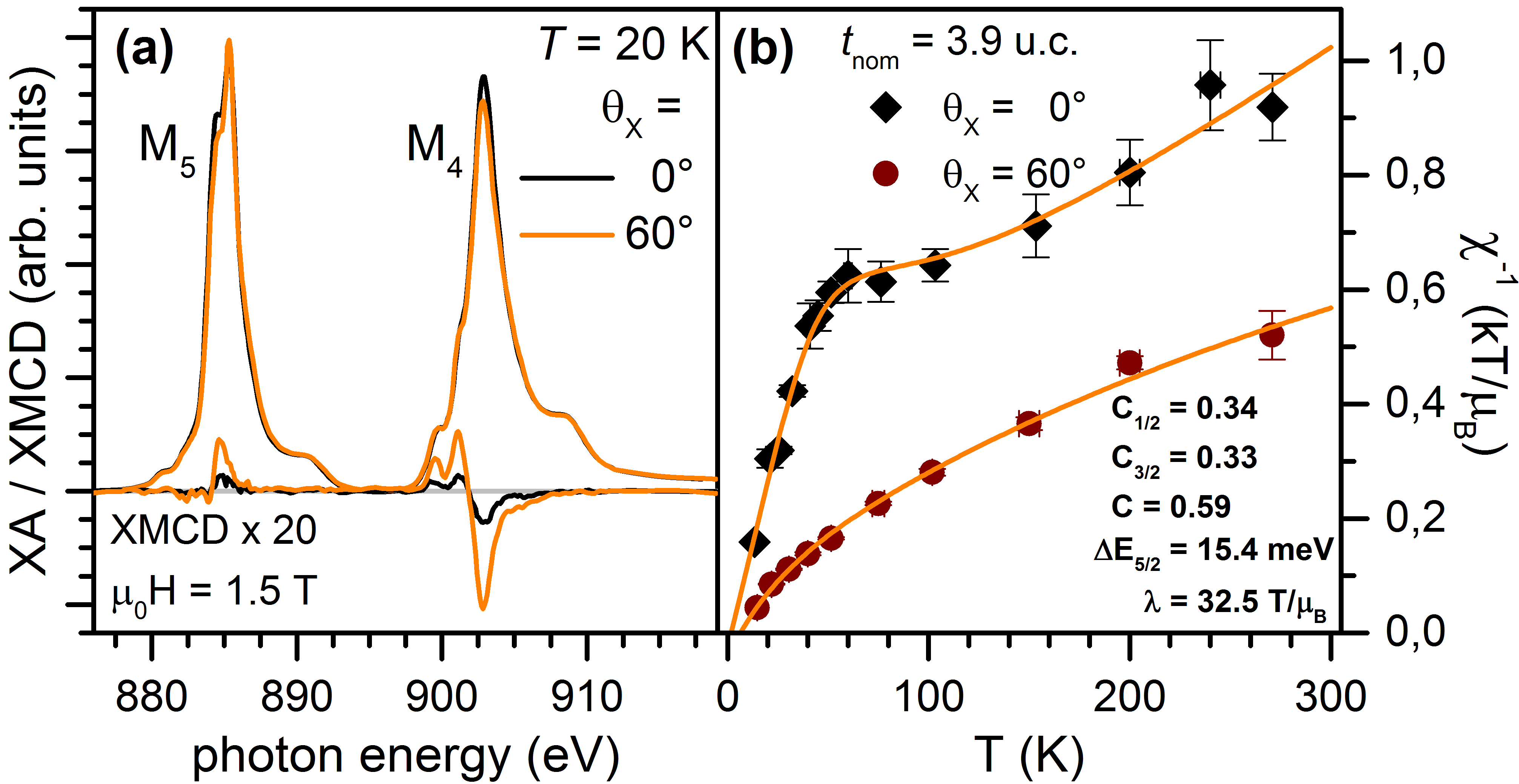} 
\caption{\label{fig4animag}
(a) XA and XMCD datasets for $\tn=4$ u.c. at normal ($\theta_X=0$) and oblique ($\theta_X=60^\circ$) incidence at $T = 20$ K and with applied field of $\mu_0H = \pm1.5$ T.
(b) anisotropic inverse susceptibility for the sample, determined from XMCD measurements such as in (a). Solid lines represent fits according to eq.~(\ref{fullmodel}). Resulting parameters are given in the legend and discussed in the text.
}
\end{figure}

The element and orbital specific measurement of the temperature dependent magnetic response provides direct access to the Ce $4f$ magnetic moments
as well as independent information on the CF splittings \cite{Prae16a}.
Figure~\ref{fig4animag}(a) displays normal and oblique incidence XA and XMCD data obtained for $\tn = 4$ u.c. at a temperature of $T=20$ K.
While the small yet finite XLD is discernible in the XA spectra, the XMCD datasets reveal a pronounced anisotropy in the paramagnetic response.
Just as  $\Upsilon>1$ determines the highest CF states to be of $|\pm5/2\rangle$ character from XLD, it follows from the strong in-plane single ion anisotropy
that $|\pm1/2\rangle$ possess large statistical weight at $T=20$ K and thus either constitute the ground state or are energetically adjacent to it.  

More insight is obtained from a quantitative evaluation of the (anisotropic) inverse susceptibility and its temperature dependence, which is shown in Fig.~\ref{fig4animag}(b).
Susceptibilities are obtained from XMCD by evaluating the magnetic Ce $4f$ polarization on the basis of the well-known sum rule for the orbital magnetic moment \cite{Thol92a}.
The direct application of the spin moment being prohibitive in case of Ce \cite{Schi94a,Tera96a,Jo97a} we proceed as in our previous work \cite{Prae15a,Prae16a} and derive the total moment by assuming  the atomic relation
$m_S = -m_L/4$ between its spin and orbital constituents to hold.
We shall neglect the possible systematic underestimation (by $\approx 13\%$) of the orbital magnetic moment from Ce M$_{4,5}$ XMCD data \cite{Prae16a},
since it constitutes only a minor correction to the effects reported below.

The experimental data of Fig.~\ref{fig4animag}(b) contain a number of characteristics to be captured by a modelling approach.
These include the reduced magnitude of the effective paramagnetic moment compared to the free ion value as well as its pronounced anisotropy.
In addition, the normal incidence data feature a strong kink at $T \approx 50$ K which results from CF splitting.
Finally, we note the occurrence of a finite, positive and anisotropic paramagnetic Curie-Weiss temperature $\Theta_p$.

The solid lines in Figure \ref{fig4animag}(b) were calculated assuming a hexagonal CF, including intersite magnetic coupling at the mean field level \cite{Prae16a}.
As mentioned above, Kondo screening is being accounted for in a phenomenological way as follows.
\begin{eqnarray} \label{anisochi}
\chi^{}_{||} &= &\frac{g^2\mu_B^2}{4k_BTZ}  \left (  C_1^2 + 9C_3^2p_1 + 25C_5^2p_2 \right ) \\ 
\chi^{}_{\perp} &= &\frac{g^2\mu_B^2}{4k_BTZ} \cdot \left (  C_1^2 \left (  9 + \frac{16k_BT}{\Delta_1} \right )+\right. \\
                 &+  &C_3^2 \left ( \frac{10k_BT}{\Delta_2-\Delta_1}-\frac{16k_BT}{\Delta_1} \right ) p_1-\left. C_5^2 \frac{10k_BT}{\Delta_2-\Delta_1}  p_2 \right )  \nonumber
\end{eqnarray}

The susceptibility at an arbitrary angle $\theta$ with respect to the hexagonal axis is then obtained via 
\begin{equation} \label{fullmodel}
\chi^{}_\theta = \frac{\cos^2\theta}{\chi_{||}^{-1}-\lambda}+\frac{\sin^2\theta}{\chi_{\perp}^{-1}-\lambda},
\end{equation}

\noindent where the interaction between Ce sites is represented by the mean field coupling constant $\lambda$.

Quite evidently, our approach is capable of  quantitatively reproducing the experimental susceptibility data.
It is in fact the approach with the smallest number of free parameters which allowed us to model the results across the entire range of CePt$_5$ thicknesses studied, and where these parameter values vary in a sensible way, largely in accordance with the respective XLD results \cite{CPdiss}.
All the many body physics is contained in the magnitudes of the $C_i$ factors, introduced such as to directly reflect the reduction of the magnetic moments in the $|\pm i/2\rangle$ states. 

In accordance with the discussion above, a small value for $\Delta_1 \approx 0.1$ eV has to be assumed in order to reproduce the anisotropy of the paramagnetic response.
As a result, the magnetic response $\chi_{||}$ along the hexagonal axis is increased in comparison with the one obtained for a pure $|\pm1/2\rangle$ Kramers doublet:
the unscreened effective moment at $\theta_X=0$ increases from $m_{\text{eff}}^{1/2}=\frac{\sqrt{3}}{2} g \mu_B \approx 0.74 \mu_B$ to $m_{\text{eff}}^{3/2}=\frac{\sqrt{15}}{2} g \mu_B \approx 1.66 \mu_B$.
Consequently, the moment reduction factors $C_{1,3} \approx 1/3$ reveal a stronger Kondo screening of the quasi-quartet ground state compared to our previous estimation in Ref.~\onlinecite{Prae15a}. 

To reproduce the kink in $\chi_{||}$, a value of $\Delta_2 = 15.4$ meV is required according to the fit. 
This value is notably smaller than the one obtained above for $\tn = 10.8$ u.c. and
$\Delta_2$ indeed exhibits a systematic dependence on the intermetallic thickness.
The Curie-Weiss behavior is accounted for by a coupling constant with $\lambda>0$ and the anisotropy of $\Theta_p$ follows naturally from the anisotropic susceptibility.

Since $\Theta_p>0$ might in principle indicate the possibility of ferromagnetic order, we have examined the low temperature behavior in greater detail for the case of $\tn = 4$ u.c.
While magnetic order is not observed down to $T\approx 2$ K, our measurements do  indicate the presence of a low temperature scale of $T^*\approx 25K$, as we shall discuss in section \ref{LT} below.

For now, we conclude our discussion of the magnetic analysis with reference to recent susceptibility calculations within the thermodynamic Bethe Ansatz \cite{Desg15a} for an impurity degeneracy of $N=4$.
This comparison is sensible only in a temperature range in which the thermal occupation of the $|\pm 5/2\rangle$ doublet may safely be neglected.
This is the case for $T\approx 30$ K , where the slope in $\chi_{||}^{-1}$ vs.~$T$ corresponds well to the moment reduction determined by $C_{1,3}$. 
On the one hand, we find from Ref.~\onlinecite{Desg15a} that in the limit of $\Delta_1 \ll T_K$ a moment reduction of this order is expected at $T/T_K \approx 0.085$, which would nicely fit with a Kondo scale in the range of some $10^2$ K.
On the other, looking at the calculated susceptibility, we find that $\chi_{||}$ should nearly have reached a temperature independent value in this temperature range, which is strongly at variance with our experimental findings.
We thus arrive at a similar conclusion as with respect to the NCA scheme above: while some useful connections with the solutions to the impurity Kondo problem can be established, the occurrence of a separate low-energy scale impedes a more thorough analysis on their basis.

\subsection{thickness dependence of XLD and XMCD}

\begin{figure} 
\includegraphics[width=\columnwidth]{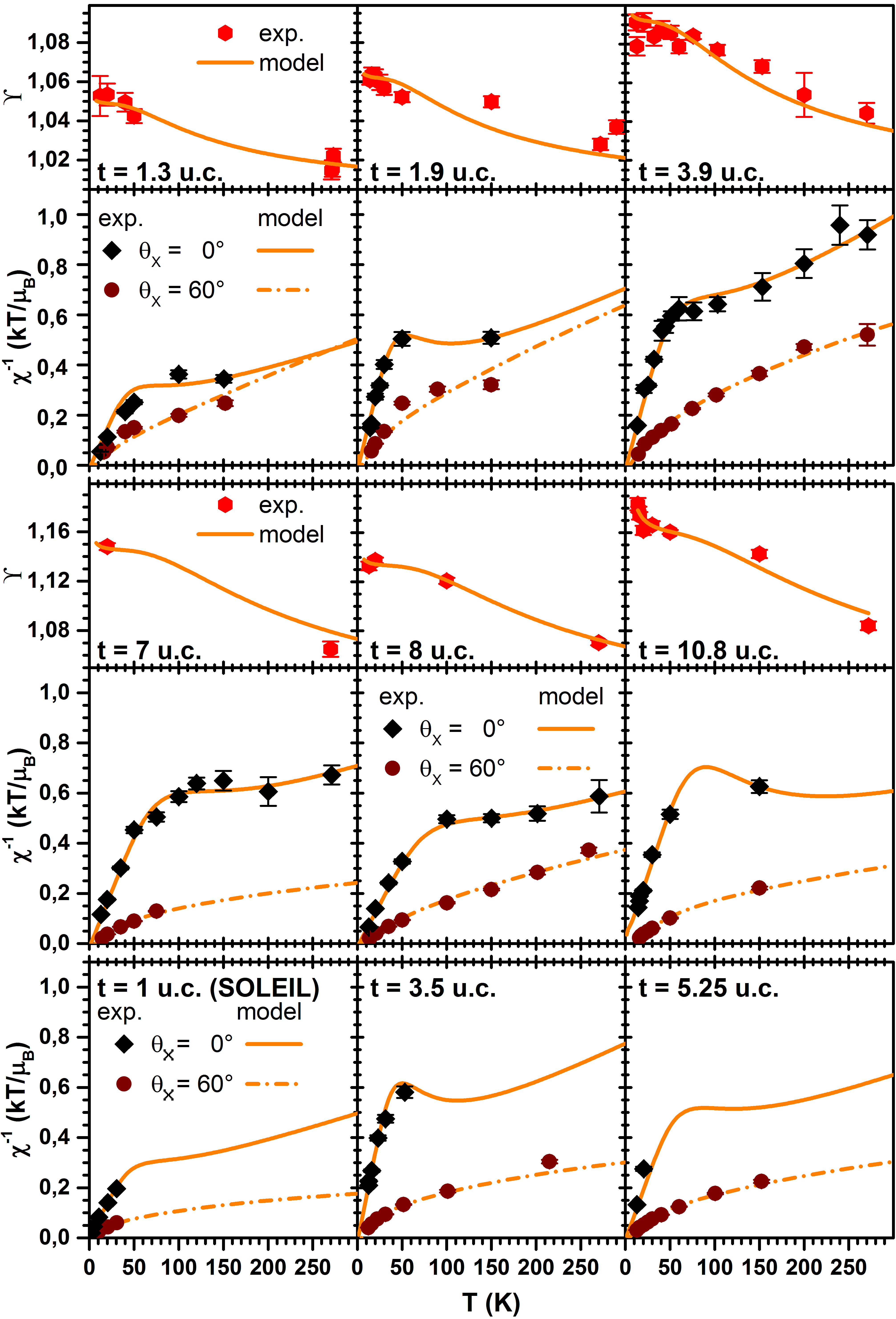} 
\caption{\label{fig5allfits}
Overview over XLD and XMCD results obtained for CePt$_5$ specimens of various thicknesses in the range 1 u.c.$\leq \tn \leq 11$ u.c.
Experimental datapoints for the XLD parameter $\Upsilon$ are shown along with the fits according to eq.~(\ref{EqUps1}),
those for the inverse susceptibility with fits according to eq.~(\ref{fullmodel}).
Where applicable, identical parameters were used in both fits.
The resulting fit parameters are given in Fig.~\ref{fig6params}.
For the sake of clarity, ordinate scales for $\Upsilon$ differ between the first and third row of panels.
Note that particularly in the latter the high temperature limiting value $\Upsilon=1$ is strongly suppressed.
}
\end{figure} 

\begin{figure} 
\includegraphics[width=\columnwidth]{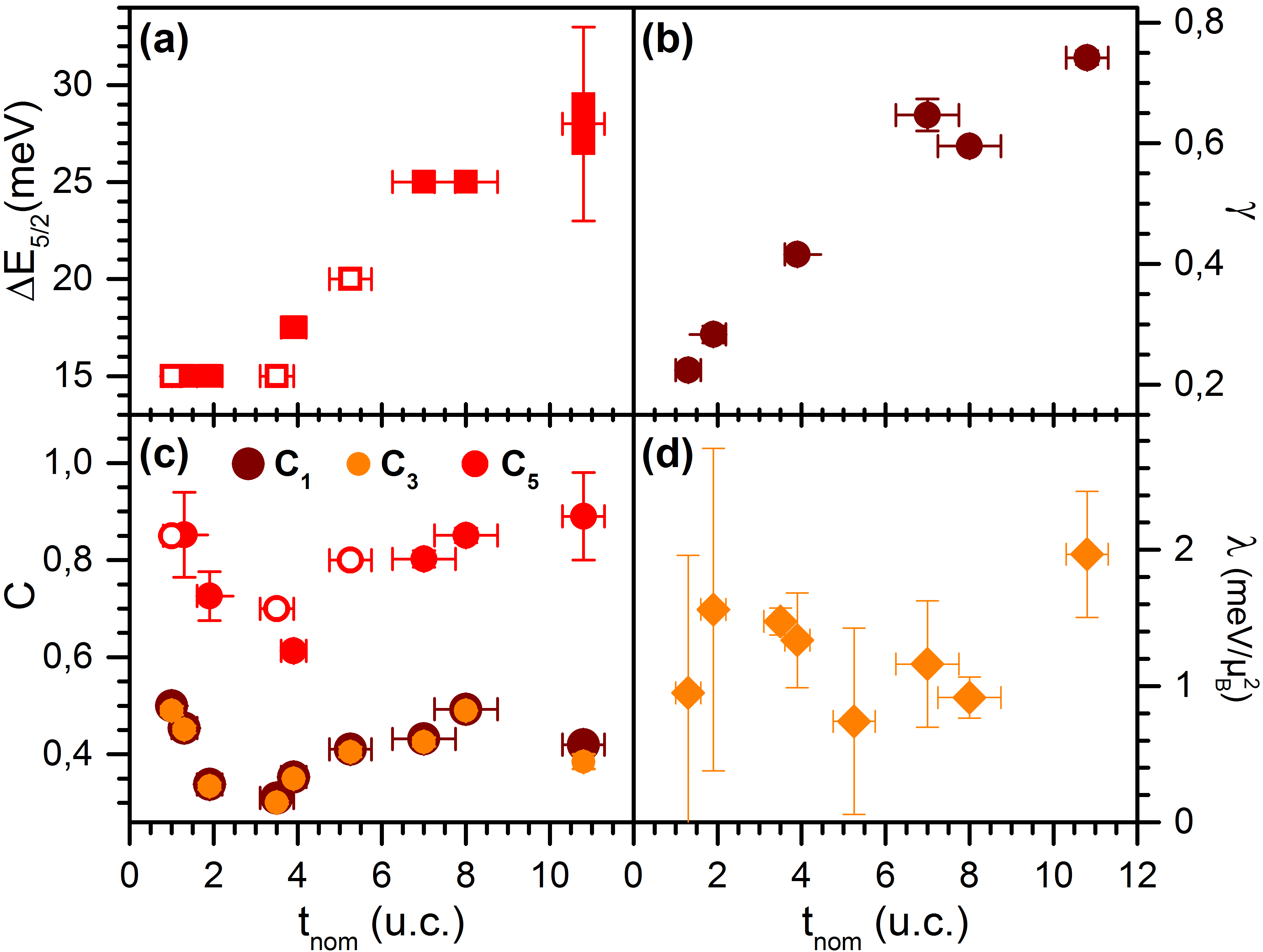} 
\caption{\label{fig6params}
Thickness dependence of the various parameters determined from least squares fitting of the experimental XLD and XMCD data in Fig.~\ref{fig5allfits}.
(a) CF excitation energy $\Delta_2$ ($\Delta_1 = 0.1$  meV assumed except for $\tn = 10.8$ u.c., see section \ref{ResXLD}).
(b) total reduction $\gamma$ of the XLD magnitude.
(c) Moment reduction factors C$_i$ as indicators of magnetic Kondo screening.
Kondo screening is strongest between $2$ u.c. $\lesssim \tn \lesssim 4$ u.c., where also hybridization is strongest \cite{Prae15a}.
(d) Mean field molecular field constant $\lambda$.
}
\end{figure}

Experiments as described above were carried out for CePt$_5$/Pt(111) specimens of various thicknesses in attempt to elucidate the impact of hybridization strength on the observed behaviors.
Figures \ref{fig5allfits}  and \ref{fig6params} summarize our findings concerning both XLD and XMCD  in terms of the 
model equations (\ref{EqUps1})  and (\ref{fullmodel}) across the range of CePt$_5$ thicknesses studied.
Fig.~\ref{fig5allfits} displays the experimental data, i.e. $\Upsilon (T)$ (where available) and the anisotropic paramagnetic response determined at $\theta_X=0^\circ$ and $\theta_X=60^\circ$.

All experimental datasets were subjected to simultaneous modelling of $\Upsilon$, $\chi_{||}$ and $\chi_{60^\circ}$.
Except for the case of $\tn = 10.8$ u.c., where $\Delta_1$ could be determined from the data, we have adopted $\Delta_1 = 0.1$ eV from $\tn = 3.9$ u.c. throughout.
With $\Delta_1$ fixed, we are left with a total of six further parameters to be determined.
Besides $\Delta_2$, which is shared by the equations for XLD and XMCD, we determine the strength of XLD reduction ($\gamma$), which comprises both effects induced by hybridization, i.e.~the altered XA line shape and the mixing of $|m_j\rangle$ weights. 
The remaining parameters apply to the magnetic data only and consist of the $C_i$ moment reduction factors and the mean field coupling $\lambda$.
For the three specimens represented in the bottom row of Fig.~\ref{fig5allfits} the values of $\Delta_2$ and $C_5$ were supplied by hand such to be in accordance with the adjacent specimens, since the amount of experimental data does not warrant their independent determination.
These cases are represented by open symbols in Fig.~\ref{fig6params} below.

But for $\chi_{60^\circ}^{-1}$ at $\tn=1.3$ u.c. and $\tn=1.9$ u.c. our modelling provides a quite satisfactory overall description of the experimental material.
The physical picture arising from these fits is best discussed by inspecting the trends of the parameter values when plotted vs.~intermetallic thickness.


A synopsis of these parameter evolutions is provided in Figure~\ref{fig6params}.
It reveals a number of systematic variations.
There is an obvious transition in the magnitude of the CF splitting $\Delta_2$. 
For the thicker, more weakly hybridized films $\Delta_2$ takes on values of $\Delta_2 \approx 25\dots 30$ meV,
considerably smaller than the estimate from bulk susceptibility measurements on polycrystalline CePt$_5$ \cite{Luek79a}.
$\Delta_2$ determines both the rate of approach of the XLD parameter $\Upsilon$ towards unity towards high temperature as well as the position of the marked kink in $\chi^{-1}$ at normal incidence.
Also, the general trend of the paramagnetic (single ion) anisotropy becoming less anisotropic as $\tn$ is reduced is compatible with a reduction in $\Delta_2$ (given that $\Delta_1$ is already small).
It is not evident to  unequivocally identify the cause of this transition.
On the one hand it seems unlikely that the variations in lattice parameter play a major role, since the structural changes (see Ref.~\onlinecite{Prae15a}) are small for $\tn \gtrsim 4$ u.c., where the transition in $\Delta_2$ is essentially taking place.
On the other, while the region of small $\Delta_2$ coincides with the occurrence of strong hybridization, there is no obvious further correlation with the nonmonotonous variation in hybridization strength vs.~$\tn$ which occurs in this range of intermetallic thickness.

The latter has  a direct bearing on the magnitude of the magnetic response, however, as is manifest from the behavior of the $C_i$ factors representing the moment reductions in Fig.~\ref{fig6params}(c).
Kondo screening is thus strongest where hybridization is strongest. 
We note that the screening factors $C_1$ and $C_3$ assume nearly identical values in the fits for all specimens.
We take this as further evidence that indeed a quasi quartet CF ground state is formed.

Another remarkable result of our experiments is the strong reduction of XLD towards small $\tn$, i.~e.~when hybridization is strong.
This reduction is given by the parameter $\gamma$ and plotted in Fig.~\ref{fig6params}(b).
The overall XLD reduction notably is much stronger than what would be expected from the asymmetric XA line shape broadening 
and thus is obviously dominated by nonthermal $|m_j\rangle$ mixing as discussed above in section \ref{general}.

With respect to Fig.~\ref{fig6params}(d) we note that a small but finite mean field coupling constant ($\lambda$) is consistently found for all specimens.
It appears, therefore, that the dominant magnetic correlations in the local moment regime of CePt$_5$/Pt(111) are ferromagnetic in nature, while  bulk CePt$_5$ orders antiferromagnetically at very low temperature \cite{Schr88a}.

The models implemented in eqns.~(\ref{EqUps1}) and (\ref{fullmodel}) thus provide a good basis for a systematic analysis of the trends generated by varying the strength of Ce $4f$ hybridization by choice of intermetallic thickness.
Nevertheless, quantitative parameter values resulting from our fits should in principle be taken with some caution. 
Kondo screening for example is inherently temperature dependent while the parameters of our model equations are not.
The fitting procedure will thus produce parameter values which best emulate this thermal behavior.
It is gratifying therefore to note that most recent experiments employing electronic Raman scattering lend strong support to our present conclusions with respect to the CF level structure \cite{HalbNNa}.

\subsection{low temperature behavior} \label{LT}

\begin{figure} 
\includegraphics[width=\columnwidth]{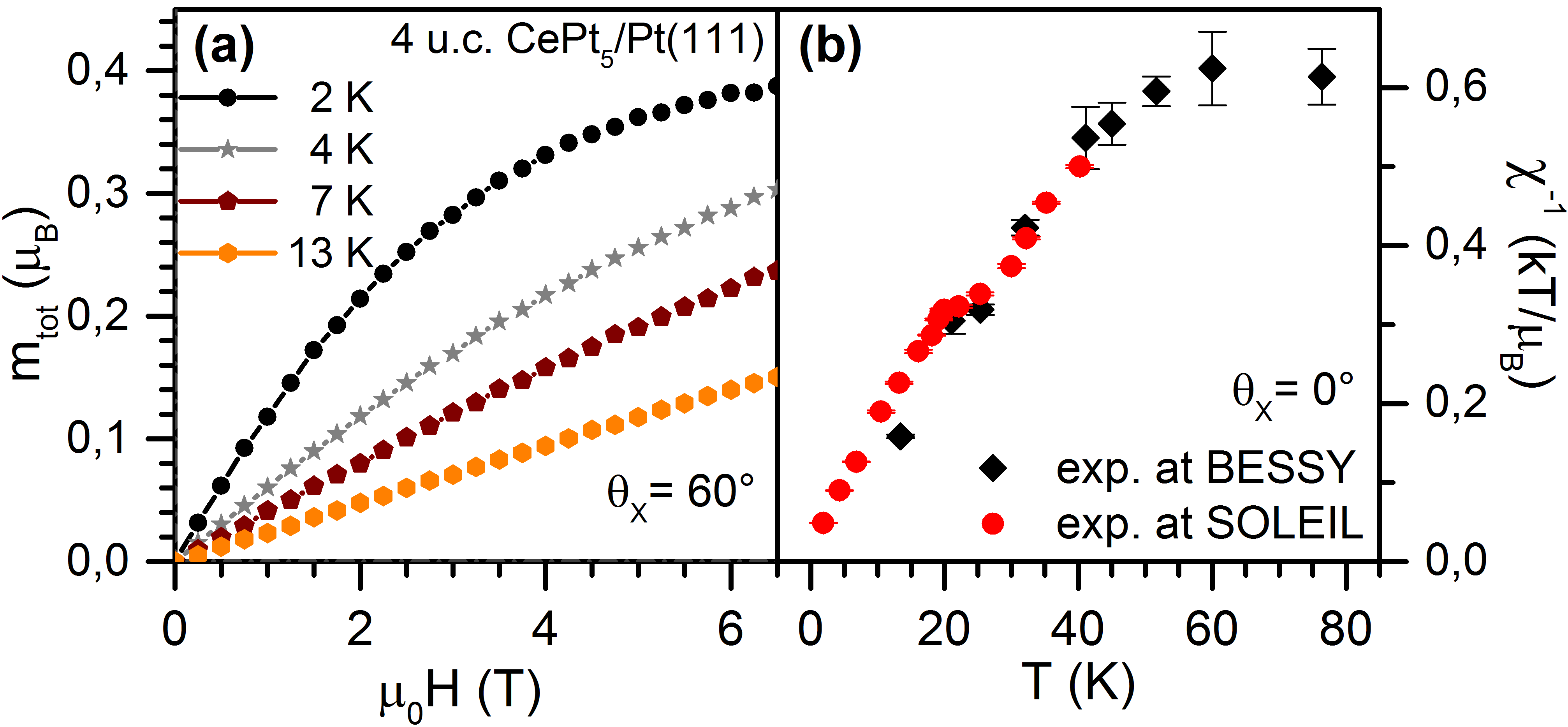} 
\caption{\label{figanomaly}
(a) Low temperature Ce $4f$ XMCD magnetization curves measured at oblique incidence for a CePt$_5$ thickness of $\tn = 4$ u.c.
(b) Detailed temperature dependence of the inverse Ce $4f$ susceptibility at NI, revealing a departure from the high temperature Curie-Weiss behavior near $T^* \approx 25$ K.
Datasets produced at BESSY and SOLEIL, respectively, yield very good agreement in the overlapping temperature range.
}
\end{figure} 

Our consistent finding of a mean field coupling constant ($\lambda > 0$) indicates the possibility of a ferromagnetically ordered ground state at temperatures below, say, $T= 5\dots10$ K.
We have therefore tested this possibility by a more detailed study of the low temperature magnetic response at SOLEIL for a specimen with $\tn = 4$ u.c.
Figure \ref{figanomaly}(a) displays a selection of XMCD magnetization curves for the lowest temperatures, measured at $\theta_X=60^\circ$, for which $\Theta_p \approx 6.5$ K according to the data of Fig.~\ref{fig4animag}.
The magnetization curves reveal a purely paramagnetic response with no sign of ferromagnetism.
Clearly, the character of magnetic response must exhibit some departure from a straightforward Curie-Weiss behavior at some intermediate temperature.

When plotting $\chi^{-1}_{||}$ vs.~temperature as in Fig.~\ref{figanomaly}(b),  such a deviation is indeed observed around  $T\approx 25$ K. 
Datasets acquired at BESSY and SOLEIL, respectively, agree very well \cite{tempernote} and the anomaly at $20\dots25$ K is in fact already present in the BESSY dataset.

The kink in $\chi^{-1}_{||}(T)$ may be seen as separating two distinct Curie Weiss regimes with $\Theta_p >0$ for $T \gtrsim 25$ K and $\Theta_p \approx 0$ for $T\lesssim 20$ K.
From extensive simulations we rule out that the observed behavior could be obtained by assuming a more refined (e.g. spatially inhomogeneous) crystal field scheme 
Also, the absence of any peculiarity in the XLD data at a temperature scale of $\approx 25$ K speaks against a CF related effect.

Instead, we notice that a departure from Curie Weiss behavior may hint at emerging lattice coherence.
This was e.~g.~also suggested \cite{yang08a} in case of CePb$_3$ which exhibits a magnetic anomaly very much reminiscent of the one observed here\cite{Duer86a}.
A coherence temperature of the order of $T^*=25$ K in the 4 u.c.~CePt$_5$ intermetallic is largely in line with the photoemission results by Klein et al.~(Ref.~\onlinecite{Klei11a}).
Obviously, the then expected crossover to the low temperature scaling regime with temperature independent Pauli susceptibility is not yet fully undergone at $T=2$ K, indicating a small degeneracy temperature $T_0 \ll T^*$ -- not infrequent in heavy electron systems \cite{Faze99a}.

Calculations within the Kondo lattice model (KLM) indicate a considerable robustness of the heavy fermion bands against temperature and magnetic fields  \cite{Beac08a,Berc12a}.
Motivated by this observation, we consider 
the implications of the experimental $M(H)$ behavior at lowest experimental temperature in Fig.~\ref{figanomaly}(a) in the framework of heavy,  itinerant Ce $4f$ states.
Such an attempt also seems worthwhile since analyzing the measured magnetization curves  in terms of local moment magnetization functions results in physically inconsistent parameters.

One characteristic experimental feature is that the magnetization curve visibly approaches some saturation behavior with a Ce $4f$ saturation magnetization of $0.4\dots0.45$ $\mu_B$ per Ce atom.
This value amounts to only a fraction of the expected saturation moment of $1.9$ $\mu_B$ per atom in the local moment picture, given the hexagonal CF scheme determined above and $\theta_X=60^\circ$.

With itinerant $4f$ electrons, the rationale for observing paramagnetic saturation is different from the case of local moments: it is expected to occur as a consequence of a Lifshitz transition induced by the applied magnetic field, i.~e.~when the Zeeman splitting shifts the chemical potential into the hybridization gap of the majority states \cite{Beac08a,Berc12a}.
Put differently, at very low temperature the scale on which magnetic saturation is observed is given by equating Zeeman and Fermi energies of the heavy band.
The magnitude of saturation magnetization then depends essentially on the fraction of the Brillouin zone covered by the heavy band and may indeed be small.

From this perspective, the Ce $4f$ magnetization approaching magnetic saturation at applied fields of the order of $\mu_0 H = 6$ T implies a heavy band with a Fermi energy around $0.6$ meV, corresponding to a degeneracy temperature of $T_F \approx 7$ K.
This is comparable to the width of the Fermi-Dirac distribution at 2 K and it follows immediately that a temperature independent Ce $4f$ contribution to the Pauli susceptibility will only be obtained at temperatures well below the range accessible to our experiments.
The KLM calculations reported in Ref.~\onlinecite{Berc12a}, when evaluated for the temperature dependent $4f$ susceptibility, appear to lend support to such an interpretation of our findings.
These caclulations do cover a parameter range down to $T \lesssim T_F \ll T^*$ and a small moment Curie Weiss like magnetic response is indeed found on this temperature scale \cite{BercxPriv}.

These considerations leave the question untouched whether additionally some kind of `two liquid' scenario might apply in analogy to those cases with $T_K < T^*$ for which this phenomenology was introduced \cite{yang08a,shir12a,yang14a,yang16a}.
Further work shall be required to more firmly establish the validity of the heavy fermion scenario to account for the Ce $4f$ magnetic response in CePt$_5$/Pt(111) and is currently in progress.

We finally emphasize that no changes of XA line shape and $4f$ occupation occur in the vicinity of the temperature of the magnetic anomaly \cite{Prae15a}.
This finding illustrates that the main role of the `delocalization process'  at $T^*$ consists of establishing phase coherence between the Ce sites, the `local physics'  remaining essentially unaltered.

\section{summary and conclusions}

In conclusion, we have presented a detailed investigation of the spectral and magnetic response as detected by x-ray absorption and dichroism at the Ce M$_{4,5}$ edges of CePt$_5$/Pt(111) ordered surface intermetallics.
Combining pieces of evidence from different spectroscopic modes and geometries we were able to gather relevant information on the interactions and associated energy scales in this material.
The general picture emerging from our study is that for $T \gtrsim 30$ K we are essentially concerned with  the ``impurity regime'' featuring
substantially Kondo-screened local moments, subjected to a hexagonal crystal field and weak ferromagnetic correlations.
The CF ground state is essentially a quasi-quartet of the $|\pm1/2\rangle$ and $|\pm3/2\rangle$ states.
The $|\pm5/2\rangle$ states are split off by $\approx 15$ meV in the range of small intermetallic thickness, whereas the splitting increases to $\gtrsim 25$ meV at larger $\tn$.

The tunability of hybridization by epitaxial strain provides us with a non-thermal control parameter in a surface science experiment.
The Ce $4f$ paramagnetic moment clearly depends on hybridization strength and we observe the strongest moment reduction in the case of strongest hybridization.
It is maybe an interesting observation that strong hybridization coincides with small overall CF splitting.
A similar correlation appears to hold in case of CeAg$_x$, albeit on a much smaller energy scale \cite{Zinnerunpub}.

While the magnitude of Kondo screening is compatible with an impurity Kondo scale of order $10^2$ K as previously determined from the temperature dependent Ce valence\cite{Prae15a},
the magnetic response at lowest temperature is not.
Instead, for specimens with $\tn = 4$ u.c. we find an anomaly in the magnetic response at $T^* = 25$ K  which we discuss as potentially signalling the onset of lattice coherence.
A coherence temperature of this order is well in line with previous experimental evidence.
Adopting this view, we may understand the occurrence of paramagnetic saturation with a small saturation moment as to emerge from a field induced Lifshitz transition.
The observation of a saturation field of the order of 6 T is then indicative of a very narrow $4f$ band with correspondingly small degeneracy temperature. 
Such a small inherent energy scale readily accounts for the fact that a temperature independent $4f$ contribution to the Pauli susceptibility is not observed within the temperature range of our experiment ($T \gtrsim 2$ K).

Altogether, our results demonstrate the promising potential of x-ray absorption measurements for investigations of correlated matter, yielding results that are complementary to other methods of surface science.
It is the unique capability of XA and XMCD to specifically detect the spectral and magnetic response of the $4f$ degrees of freedom which allows one to gain insight in several small energy scales coexisting in heavy fermion materials.
We hope that our work will stimulate interest in carrying over the methodology to other heavy fermion materials.
In particular, we anticipate that the element and orbital specificity of XLD and XMCD should provide profound insight into the physics behind metamagnetic transitions in those heavy fermion materials where the required magnetic field is accessible with current synchrotron radiation instrumentation.

\begin{acknowledgments}
Acknowledgement for assistance on the occasion of various synchrotron radiation beam times is owed to M.~Zinner, H. Kie\ss ling, B. Muenzing, P.~Sprau, and S. Br{\"{u}}ck
 as well as to the beamline staff, T. Kachel (HZB), F. Choueikani and P. Ohresser (SOLEIL) for their support. 
We also thank P.~Hansmann, M.~W.~Haverkort, F.~F.~Assaad, M.~Bercx, H.~Schwab and F.~Reinert for most helpful and stimulating discussions. 
This work received financial support by the Deutsche Forschungsgemeinschaft within FOR1162 (TP 7).
Access to synchrotron radiation was also partially granted by HZB managed funds and the European Community's Seventh Framework Programme (FP7/2007-2013) under the CALIPSO project (Grant Agreement No. 226716).
Generous allocation of beam time at the synchrotron radiation facilities as well as their general support is gratefully acknowledged.
\end{acknowledgments}

\appendix*
\section{Modeling the XLD parameter $\Upsilon$} \label{Upsilon}

The reduction of both the TEY signal and the relative magnitude of XLD with decreasing CePt$_5$ thickness prompted a search for an alternative, simple and robust measure of XLD, suitable for a determination of the CF splittings $\Delta_{1,2}$.
It turns out that by using the relative peak heights of the spectral features B and C at the M$_5$ and M$_4$ edges of NI spectra, a suitable parameter $\Upsilon$ can be obtained.
Its sensitivity for the magnitude of XLD relates to the fact that the XLD of features B and C possess opposite sign for all $m_j$.
The relation to XLD is obtained by relating this peak ratio to the one obtained at oblique incidence (i.e.~the isotropic spectrum in case of the data taken at BESSY II).
We define $\Upsilon$ as

\begin{equation} \label{DefUps}
\Upsilon(T) =\frac{ I^{NI}_{\text{B}}  }{I^{NI}_{\text{C}} } \cdot \frac{ I^{ISO}_{\text{C}}}{I^{ISO}_{\text{B}}}=\frac{ I^{NI}_{\text{B}}  }{I^{ISO}_{\text{B}} } \cdot \frac{ I^{ISO}_{\text{C}}}{I^{NI}_{\text{C}}}.
\end{equation}

In practice, $\Upsilon$ is most readily evaluated by direct comparison of the peak intensities between normal and oblique incidence spectra at the spectral positions of features B and C,
as indicated in the regrouped, final expression of eq.~(\ref{DefUps}).

\begin{figure}
\includegraphics[width=\columnwidth]{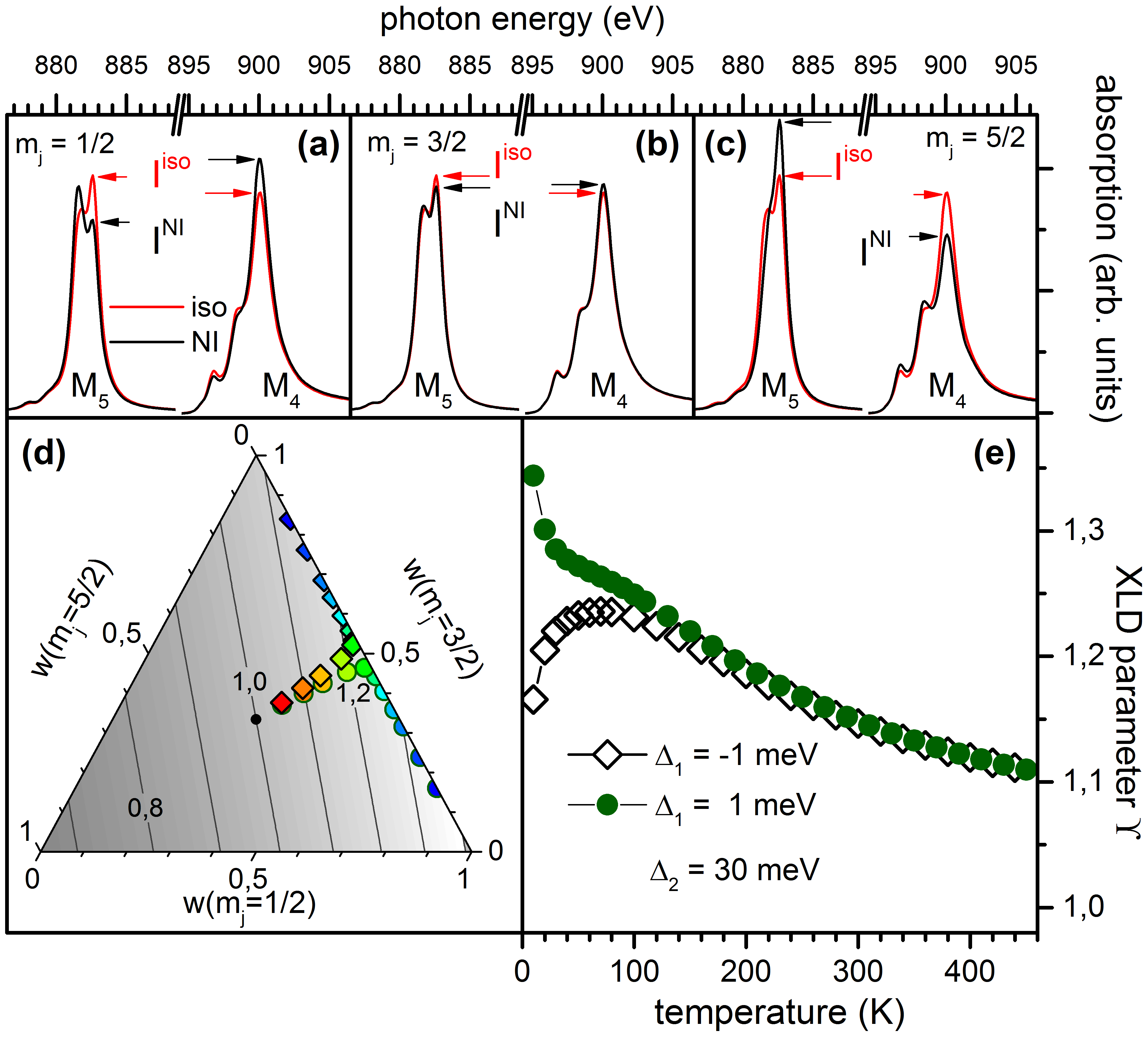}
\caption{\label{figUpsilon}
(a-c) calculated normal incidence and isotropic spectrum for each of the $m_j$ states.
Arrows indicate the intensities utilized for determining $\Upsilon$ (eq.~(\ref{DefUps})). 
(d) ternary diagram representing the dependence of $\Upsilon$ on the fractional occupations $w(m_j)$. 
Colored symbols represent the trajectories of $\Upsilon$ followed for $\Delta_1=\pm1$ meV, $\Delta_2=30$ meV.
Colors represent temperature and range  from red ($T= 900$ K) to blue, $T= 7$ K).
(e) temperature dependence $\Upsilon(T)$ for the same $\Delta_{1,2}$.
}
\end{figure}

At a given temperature $T$, the magnitude of $\Upsilon$ will depend on the occupation of the CF levels, each contributing in an individual way.
As an illustration, we plot the NI spectra for each of the CF states along with the isotropic spectrum in panels (a)-(c) of Fig.~\ref{figUpsilon}, calculated here without taking asymmetric broadening into account.
The relevant intensities at features B and C are indicated by arrows.

For each of the $|\pm i/2\rangle$ doublets, we determine the deviation of the ratios $I^{NI}_{i,\text{B}} / I^{ISO}_{i,\text{B}}$ and $I^{NI}_{i,\text{C}} / I^{ISO}_{i,\text{C}}$ from unity

\begin{equation} \label{Afactors}
A_{i,B} = \frac{I^{NI}_{i,\text{B}}}{ I^{ISO}_{i,\text{B}}}-1 , \,\,\,\,
A_{i,C} = \frac{I^{NI}_{i,\text{C}}}{ I^{ISO}_{i,\text{C}}}-1
\end{equation}

\noindent from the calculated multiplet spectra and express $\Upsilon(T)$ in these quantities as follows

\begin{equation} \label{EqUps1}
\Upsilon(T) = 1 + \gamma \left (
\frac{1+
\left ( A_{1,B} + p_1 A_{3,B} + p_2 A_{5,B}   \right ) / Z}
       {1+
\left ( A_{1,C} + p_1 A_{3,C} + p_2 A_{5,C}   \right ) / Z}
-1 \right ),
\end{equation}

\noindent with $p_{1,2}$ and $Z$ as defined in section \ref{ResXLD}.
 In this expression, $\gamma$ represents an overall reduction of the XLD magnitude, which comprises both its reduction due to the asymmetric spectral response and  the hybridization induced mixing of $|m_j\rangle$ states.
The temperature dependence $\Upsilon(T)$ is then encoded in the Boltzmann weights $p_{1,2}(T)$.

To more accurately represent the fact that XLD reduction due to asymmetric spectral response is different for features B and C (see Fig.~\ref{fig2XASXLD}(c) and corresponding text)
one can either determine the quantities defined in eq.~(\ref{Afactors}) from calculated spectra taking the spectral asymmetry into account or by introducing separate XLD reduction factors $\gamma^{\,}_{B}$ and $\gamma^{\,}_{C}$.

\begin{equation} \label{EqUps2}
\Upsilon(T) =
\frac{1+\gamma^{\,}_{B}\left ( A_{1,B} + p_1 A_{3,B} + p_2 A_{5,B}   \right ) / Z}
       {1+\gamma^{\,}_{C}\left ( A_{1,C} + p_1 A_{3,C} + p_2 A_{5,C}   \right ) / Z}
\end{equation}

Likewise, one might wish to represent the possibility that since $\Delta_2 \gtrsim T_K$ the XLD could actually be less strongly reduced for the $|\pm 5/2 \rangle$ states.
Adding such details to the model, however, does not significantly alter the fit results concerning the magnitudes of $\Delta_{1,2}$, which we are primarily interested in here.
The $\Upsilon(T)$ calculations represented in Fig.~\ref{fig3XLD11uc}(b) and Fig.~\ref{fig5allfits} were thus all computed according to eq.~(\ref{EqUps1}).
The overall XLD reduction factor $\gamma$ resulting from the fits is reported in Fig.~\ref{fig6params}(b).

Figure \ref{figUpsilon}(d) contains a ternary nomogram which represents the behavior of $\Upsilon$ as a function of the statistical weights of the $|m_j\rangle$ states.
The center of the triangle corresponds to the high temperature limit in which all $|m_j\rangle$ states possess the same weight and where therefore $\Upsilon=1$.
Lines represent initial state compositions of equal $\Upsilon$.
It is readily seen that $\Upsilon$ is primarily sensitive to the degree of admixture of $|5/2\rangle$ character and hence well suited to determine $\Delta_2$.
$\Upsilon>1$ at all temperatures, as observed in the CePt$_5$ specimens, signifies that $|\pm5/2\rangle$ is the CF state of highest energy.

Panel (e) of Fig.~\ref{figUpsilon} displays $\Upsilon(T)$, as calculated using eq.~(\ref{EqUps1}) with $\gamma=1$ (or, equivalently,  eq.~(\ref{EqUps2}) with $\gamma_B=\gamma_C=1$), setting $\Delta_2=30$meV  and $\Delta_1=\pm 1$ meV, respectively.
The high temperature behavior is dominated by $\Delta_2$ and thus similar in both cases, while the low temperature trends in $\Upsilon(T)$ are determined by the sign of $\Delta_1$.


\begin{thebibliography}{68}
\bibitem [{tem()}]{tempernote}%
  \BibitemOpen
  \href@noop {} {}\bibinfo {note} {The lowest temperature data point of the
  BESSY appears to disagree with the data gathered at SOLEIL. We ascribe this
  difference mostly to the construction details of the respective cryostats and
  positioning of temperature sensors in particular. For the SOLEIL dataset,
  toward low temperature, sensor reading is a lower limit to the actual sample
  temperature, the deviation being estimated to $\Delta T \approx 0.5$ K at
  $T_{\text{cryo}}= 2$ K. In the cryostat used at BESSY, the value read from
  the detector is rather an upper limit to the specimen temperature, some
  deviation arising for $T\lesssim 20$ K.}\BibitemShut {Stop}%
\end{thebibliography}

%

\end{document}